\documentclass[11pt,a4paper]{article}
\usepackage{jheppub}

\title{The ``universal property''  of Horizon Entropy Sum of Black Holes in Four Dimensional Asymptotical (anti-)de-Sitter Spacetime Background}

\author[a]{Jia Wang,}
\author[a]{Wei Xu,}
\author[a,b]{Xin-he Meng}

\affiliation[a]{School of Physics, Nankai University, Tianjin 300071, China}
\affiliation[b]{Kavli Institute of Theoretical Physics China, CAS, Beijing 100190, China}

\emailAdd{wangjia2010@mail.nankai.edu.cn}
\emailAdd{xuweifuture@mail.nankai.edu.cn}
\emailAdd{xhm@nankai.edu.cn}

\abstract{We present a new ``universal property''  of entropy, that is the
``entropy sum'' relation of black holes in four dimensional
(anti-)de-Sitter asymptotical background. They depend only on the
cosmological constant with the necessary effect of the un-physical
``virtual'' horizon included in the spacetime where only the
cosmological constant, mass of black hole, rotation parameter and
Maxwell field exist. When there is more extra matter field in the
spacetime, one will find the ``entropy sum''  is also dependent of
the strength of these extra matter field. For both cases, we
conclude that the ``entropy sum''  does not depend on the conserved
charges $M$, $Q$ and $J$, while it does depend on the property of
background spacetime. We will mainly test the ``entropy sum''
relation in static, stationary black hole and some black hole with
extra matter source (scalar hair and higher curvature) in the
asymptotical (anti-)de-sitter spacetime background. Besides, we
point out a newly found counter example of the mass independence
of the ''entropy product'' relation in the spacetime with extra
scalar hair case, while the ``entropy sum'' relation still holds.
These result are indeed suggestive to some underlying microscopic
mechanism. Moreover, the cosmological constant and extra matter
field dependence of the ``entropy sum''  of all horizon seems to
reveal that ``entropy sum'' is more general as it is only related
to the background field. For the case of asymptotical flat
spacetime without any matter source, we give a note for the Kerr
black hole case in appendix. One will find only mass dependence
of ``entropy sum'' appears. It makes us believe that, considering
the dependence of ``entropy sum'', the mass background field may be
regarded as the next order of cosmological constant background
field and extra matter field. However, fully explaining the
relationship between the ``entropy sum''  relation and background
properties still requires further exploration.}


\arxivnumber{1310.6811}

\begin{document}

\maketitle
\flushbottom

\section{Introduction}
Understanding the origin of black hole entropy at the microscopic
level has been a major challenge in quantum theories of gravity in
the past years. More recent interests have been focused on the
''area product'' or ''entropy product'' of the black holes which
possess more than one horizon
\cite{Cvetic:2010mn,Castro:2012av,Detournay:2012ug,Visser:2012zi,Visser:2012wu,Chen:2012mh,Toldo:2012ec,
Faraoni:2012je,Castro:2013kea,Castro:2013pqa,Lu:2013eoa,Cvetic:2013eda,Abdolrahimi:2013cza,Lu:2013ura,Anacleto:2013esa,
Chow:2013tia}. It seems clear that this additional thermodynamic
relation of entropy appears to be ''universal'' and may provide
further insight into the quantum physics of black holes. Various
black holes include Einstein-Maxwell gravity and in Super-gravity
models are tested
\cite{Cvetic:2010mn,Visser:2012wu,Toldo:2012ec,Lu:2013eoa,Cvetic:2013eda,Abdolrahimi:2013cza,Lu:2013ura,Chow:2013tia}.
The product of entropy is once expected to be more universal and
in fact independent of the mass of the black hole
\cite{Cvetic:2010mn,Castro:2012av,Chen:2012mh,Visser:2012zi,Toldo:2012ec,Castro:2013kea,Lu:2013eoa,Cvetic:2013eda,Abdolrahimi:2013cza,Lu:2013ura,
Anacleto:2013esa,Chow:2013tia}. However, It fails in some cases
\cite{Detournay:2012ug,Visser:2012wu,Faraoni:2012je,Castro:2013pqa}.
For example, in discussing the Schwarzschild-de Sitter black hole
and Reissner-Nordstrom-anti-de Sitter black hole in $3+1$
dimensions, it has been shown that the product of event horizon
area and cosmological horizon area is not mass independent, even
if including the effect of the third un-physical ``virtual''
horizon the result does not improve  \cite{Visser:2012wu}. This
mass-dependence of the product of physical horizon areas is soon
discussed more clearly in the higher curvature gravity models
\cite{Castro:2013pqa}.

Now people expect more general additional thermodynamic relation.
It is the primary work of this paper. In our present research, we
find another ``universal property''  of entropy, the ``entropy sum''
relation of black holes in four dimensional (anti-)de-Sitter
asymptotical background. They depend only on the cosmological
constant with including the necessary effect of the un-physical
``virtual'' horizon in the spacetime where only cosmological
constant, mass of black hole, rotation and Maxwell field exist.
When there is more extra matter field in the spacetime, one will
find the ``entropy sum''  is also dependent of the strength of
these extra matter field. For both cases, we conclude that the new
``universal property'', that is, the``entropy sum'' relation does
not depend on the conserved charges $M$ (mass), $Q$ (charge from
Maxwell field) and $J$ (from rotation case), while it does depend
on the property of background spacetime. To express it more
accurately, it does depend on those constants, which characterize
the strength of the background fields. We will mainly test the
``entropy sum'' relation in static, stationary black hole and some
black hole with other extra matter source (scalar hair and higher
curvature terms) in asymptotical (anti-)de-sitter spacetime
background. Besides, we point out a newly discovered failed
example of the mass independence of the ``entropy product''
relation in the discussion about the spacetime with scalar hair,
while the ``entropy sum'' relation still holds. These result are
indeed suggestive of some underlying microscopically. Anyway, the
cosmological constant and extra matter field dependence of the
``entropy sum'' relation of all horizon seems to reveal that
``entropy sum'' is more general and is only related to the
background field. One may be curious about the ``entropy sum'' in
asymptotical flat spacetime without any matter source. We will
give a note of Kerr black hole case in appendix
\hyperref[appsec:Kerr solution in flat]{B}, in which one will find
only mass dependence of ``entropy sum''  appears. It makes us
believe that, considering the dependence relation of the ``entropy
sum'' , the mass background field may be regarded as the next order
of cosmological constant background field and extra matter field,
while the Maxwell field and ``rotation field'' always play no role.
Explaining the relationship between the``entropy sum'' and
background spacetime properties still are open problems and left
to be a future work.

The surprising discovery of the cosmic late stage accelerating
expansion has inspired intensive research on the universe
background cosmological constant problem, including its functioning in astrophysics. The present paper is
organized as follows: in next Section, we test the``entropy sum''
in (A)dS black hole without other extra matter field, including
the Schwarzschild-de-Sitter solution, Reissner-Nordstrom-de-Sitter
solution and the Kerr-(anti-)de-Sitter solution; Section 3 is
devoted to the discussion of ``entropy sum'' of the black holes
with extra scalar hair and the charged rotating and static black
holes in Einstein-Weyl theory. We derive the ``entropy sum''  and
the dependence of background field constant for each of these
black hole solutions. In the end of the paper, we make some
conclusion and discussion.

\section{``Entropy sum'' of (A)dS black hole with charge and rotation}
In this section, we test the ``Entropy sum''  of (A)dS black hole
without other extra matter field, including
Schwarzschild-de-Sitter, Reissner-Nordstrom-de-Sitter Solution and
Kerr-(anti-)de-Sitter solution in four dimensions. Here are only
cosmological constant, mass of black hole, rotation and Maxwell
field in the spacetime. When there is no other extra matter fields
in the (A)dS spacetime, one will find the ``entropy sum''  of black
holes depend only on the cosmological constant with the necessary
effect of the un-physical ``virtual'' horizon included. The Maxwell
field and ``rotation field'' always play no role.

\subsection{Warm-up: Schwarzschild-de-Sitter and Reissner-Nordstrom-de-Sitter Solution}
\label{subsec:Schwarzschild-de-Sitter and
Reissner-Nordstrom-de-Sitter Solution} To give a warm-up, we begin
the discussion with the simple static and uncharged example, four
dimensional Schwarzschild-de-Sitter solution which is behaviours
as
\begin{align}
ds^2=-\Delta(r)dt^2+\frac{dr^2}{\Delta(r)}+r^{2}\left(d\theta^{2}+\sin^{2}d\varphi^{2}\right),
\end{align}
where $M$ is the mass of the black hole and $\Lambda$ is the cosmological constant, and the horizon function is
\begin{equation}
\Delta(r)=1-\frac{2M}{r}-\frac{\Lambda r^2}{3}
\end{equation}
We will substitute $\Lambda=\frac{1}{L^2}$ for convenience in this subsection. As we are aim to the ``entropy sum''  of black hole horizons, we first list the three roots of $\Delta(r)$  \cite{Visser:2012wu}
\begin{align*}
r_{1} &=2L\sin\left(\frac 13 \arcsin \left(\frac{3 M}{L}\right)\right)\\
r_{2} &=2L\sin\left(\frac 13 \arcsin \left(\frac{3 M}{L}\right)+\frac{2\pi}{3}\right)\\
r_{3} &=2L\sin\left(\frac 13 \arcsin \left(\frac{3 M}{L}\right)-\frac{2\pi}{3}\right)
\end{align*}
where, $r_{1}$ represents an event horizon and $r_{2}$ is a cosmological horizon. Both are physical
horizons. The third one $r_3$, however, is not a physical, said to be a ``virtual'' horizon. It is easy to find
the product of event horizon area and cosmological horizon area is not mass independent, even including the effect of the third un-physical ``virtual'' horizon does not improve the result \cite{Visser:2012wu}.

On the other hand, there is an exact result that
\begin{align}
\sum_{i=1}^3r_{i}^{2}=6 L^{2},
\end{align}
which immediately deriving the ``area sum'' of all horizons as
\begin{align}
\sum_{i=1}^3A_{i}=24 \pi L^{2}=\frac{24 \pi}{\Lambda}.
\end{align}
We note that the sum of the areas is a constant directly related to the spacetime background
i.e. cosmological constant. In fact, for Schwarzschild-de-Sitter black hole, the entropy of each
horizon (include the ``virtual'' one) is $S_{i}=A_{i}/4=\pi r^2$. So we conclude
\begin{align}
\sum_{i=1}^3S_{i}=\frac{6 \pi}{\Lambda},
\end{align}
the ``entropy sum'' of all four horizons is only cosmological constant dependence and also mass independence.

However, there is no conserved charges in the Schwarzschild-de-Sitter spacetime. To be more convictive, we present
the Reissner-Nordstrom-de-Sitter solution as a second warm-up. The Reissner-Nordstrom-de-Sitter solution is
\begin{align}
ds^2=-\Delta(r)dt^2
     +\frac{dr^{2}}{\Delta(r)}+r^2\left(d\theta^{2}+\sin^{2}\theta
     d\varphi^{2}\right),
\end{align}
where
\begin{equation}
\Delta(r)=1-\frac{2M}{r}+\frac{Q^2}{r^2}-\frac{\Lambda r^2}{3}.
\end{equation}
In principle the quartic can be solved explicitly, but here it is not necessary to list the
roots. This argument is shown in detail in \cite{Visser:2012wu} that there are four physical roots: each of them stand for the event horizon, Cauchy horizon, cosmological horizon and an un-physical virtual horizon respectively. The mass independence of ``entropy product'' of all horizons still hold \cite{Visser:2012wu}.

For our interest, the ``area sum'' of all four horizons is
\begin{align}
&\sum_{i=1}^4r_i^2=(r_1+r_2+r_3+r_4)^2-2\sum_{i<j}r_{i}r_{j} \notag=6 L^2\\
&\sum_{i=1}^4A_i=24\pi L^2
\end{align}
Again,
\begin{align}
\sum_{i=1}^4S_{i}=\frac{6 \pi}{\Lambda},
\end{align}
the ``entropy sum'' of all four horizons is only cosmological constant dependence and also does not depend on the conserve charges: mass $M$ and charge $Q$. Namely, Mass and the Maxwell field do no effect on the ``entropy sum''.

\subsection{Kerr-(anti-)-de-Sitter Black Holes}
\label{subsec:Kerr-(anti-)-de-Sitter Black Holes}
We continue our discussion with cosmological constant, mass and angular momentum of black hole exist in the spacetime, i.e. the familiar Kerr-(anti-)de-Sitter black hole \cite{Ghezelbash:2004af,Gibbons:2004ai,Gibbons:2004uw}
\begin{align}
ds^2=-\frac{\Delta}{\rho^2}\left({\it dt}-{\frac {a{\sin}^{2}\theta d\varphi}{\Xi}}\right)^2+\frac
      {{\rho}^{2}{dr}^{2}}{\Delta}+\frac {\rho^2 d\theta^2}{\Delta_{\theta}}  \notag
     +\frac{\Delta_{{\theta}}{\sin}^{2}\theta}{\rho^2}\left(adt-\frac {\left( {a}^{2}+{r}^{2}
      \right)d\varphi }{\Xi}\right)^2,
\end{align}
where
\begin{align*}
\rho^{2}&={a}^{2}+{r}^{2}{\cos}^{2}\theta, &
\Delta&= \left( {a}^{2}+{r}^{2} \right)
 \left( 1\mp{\frac {{r}^{2}}{{l}^{2}}} \right) -2mr,\\
\Delta_{\theta}&=1\pm{\frac {{a}^{2}{\cos}^{2}\theta}{{l}^{2}}}, &
\Xi&=1\pm\frac {a^2}{l^2},
\end{align*}
and the cosmological constant $\Lambda=\pm\frac{3}{l^2}$.
Here, the upper and lower of sign stand for the dS and AdS solution respectively.
The four roots of $\Delta$ is shown in \cite{Ghezelbash:2004af}. They satisfy the following equality
\begin{align}
&\sum_{i=1}^4 r_i^2=2l^2-2a^2, \quad\text{for dS spacetime;}\\
&\sum_{i=1}^4 r_i^2=-2l^2-2a^2, \quad\text{for AdS spacetime;}
\end{align}
It is well known that the area for each horizon is
\begin{equation}
A(r_i)=\frac {4\pi(r_i^2+a^2)}{1\pm\frac {a^2}{l^2}}
\end{equation}
Thus we obtain the ``area sum'' of all four horizons
\begin{align}
\sum_{i=1}^4 A(r_i)=\pm8\pi l^2=\frac{24\pi}{\Lambda}
\end{align}
with the ``entropy sum''
\begin{equation}
\sum_{i=1}^4 S(r_i)=\frac{6\pi}{\Lambda}.
\label{kerr}
\end{equation}
Then we demonstrate that, in four dimensional (A)dS spacetime, the cosmological constant dependence of ``entropy sum'' of all horizons is a universal property. The ``entropy sum'' is the constant, which is proportional to cosmological radius and inversely proportional to cosmological constant, no matter charge and rotation exist in the spacetime. That is to say,  mass, the Maxwell field and ``rotation field'' do no effect on the ``entropy sum''.

\section{``Entropy sum'' of (A)dS black hole with other extra matter field}
This section is devoted to the discussion about the ``Entropy sum''  of four dimensional (A)dS black hole with other extra matter field. The extra matter field of example we present here is the scalar field and with higher curvature terms. When there is scalar field in the spacetime, one will find that the ``entropy sum''  is dependent of the cosmological constant and the constant signifying the strength of self-interacting potential for the scalar field in both the conformally coupling frame and the minimally coupling frame. When we consider the charged rotating and static black holes in Einstein-Weyl theory, the ``entropy sum''  is shown a only dependence of the constant characterizing the strength of higher curvature terms, even the cosmological constant dependence is vanishing. For both cases, we conclude that the new ``universal property'', ``entropy sum''  does not depend on the conserved charges $M$ (mass), $Q$ (from Maxwell field) and $J$ (from rotation). One can believe that, ``entropy sum''  of all horizons (including the ``virtual'' horizon) does depend on those constants, which characterize the strength of the background fields.

\subsection{Scalar Hairy Black Holes}
\label{subsec:Scalar Hairy Black Holes}
We consider the Einstein-Maxwell system in four dimensions with a cosmological constant $\Lambda$ and a real conformally coupled self-interacting scalar field, described by the action
\begin{align}
\mathcal{L}=\int d^4x\sqrt{-g}\left(\frac{R-2\Lambda}{16\pi}-\frac 12
g^{\mu\nu}\partial_{\mu}\phi\partial_{\nu}\phi-\frac{1}{12}R\phi^2-\alpha\phi^4\right)-\frac{1}{16\pi}
\int d^4x\sqrt{-g}F^{\mu\nu}F_{\mu\nu},
\end{align}
where the parameter $\alpha$ is arbitrary self-interaction constant, which signify the coupling strength between gravity and the scalar field. The first well-known solution for this action is the ``MTZ'' black hole \cite{Martinez:2002ru}. Here we will focus on the charged ``MTZ'' black hole solution with the metric
\begin{align}  \label{eq:Scalar Hairy BH metric}
ds^2=-f(r)dt^2+
     \frac{dr^2}{f(r)}
     +r^2d\Omega^2
\end{align}
where $d\Omega^2$ is the line element of the 2-dimensional surface $\Sigma$
\begin{equation}
d\Omega^2=\begin{cases}
d\theta^{2}+\sin^{2}\theta d\varphi^{2},& \text{sphere } S^2;\\
d\theta^{2}+\theta^2 d\varphi^{2},& \text{flat } \mathbb{R}^2;\\
d\theta^{2}+\sinh^{2}\theta d\varphi^{2},& \text{hyperbolic } H^2.
\end{cases}
\end{equation}
The metric function is
\begin{align}  \label{f(r)}
  f(r)=-\frac{\Lambda r^2}{3}+\gamma\left(1+\frac{\mu}{r}\right)^2
\end{align}
with the electromagnetic potential is given by
\begin{equation}
A=-\frac qr dt
\end{equation}
The parameter $\gamma$ denotes
the normalized curvature constant of the 2-dimensional sub-manifold $\Sigma$. $\gamma$ can take
$+1$,$0$ and $-1$, corresponding to sphere $S^2$, flat $\mathbb{R}^2$ and hyperbolic manifold $H^2$. We will only consider $\gamma=\pm1$, as the flat case corresponds to no black hole but naked singularity.
The integration constants $\mu$ and $q$ are not independent since they must satisfy
\begin{align}
  q^2=\gamma G\mu^2\left(1+\frac{2\pi \Lambda G}{9\alpha}\right)
  \label{muq}
\end{align}
We will set $G=1$ in what follows. On the other hand, $\mu$ and $q$ are relate to the conserved charges: mass $M$ and electric charge $Q$:
\begin{align}
M=-\gamma\frac{\sigma}{4\pi}\mu \qquad Q=\frac{\sigma}{4\pi}q
\end{align}
Where $\sigma$ is the ''unit'' area of the 2-dimensional surface $\Sigma$
\begin{equation}
\sigma=\begin{cases}
4\pi,& \gamma=1;\\
4\pi(g-1),g\geq2,\quad g\text{ is the genus},& \gamma=-1.
\end{cases}
\end{equation}

However, Eq.(\ref{muq}) shows that $M$ and $Q$ are not independent in this spacetime. Thus, one can expect that the ``entropy product'' of all horizons is dependent on the conserved charges $Q$, which means a mass dependence as well. This a new failed example of the mass independence of ''entropy product'' relation.

There are two parameters $\Lambda=\pm\frac{3}{l^2}$ and $\gamma=\pm1$ in the solution (\ref{f(r)}). They will corresponds to four solutions of the action. To give a brief check of the ''entropy sum'' relation, we will only present the discussions of two of these four solutions here. We focus on the dS black hole with sphere horizon ($\Lambda=\frac{3}{l^2}$ and $\gamma=1$) \cite{Barlow:2005yd} and the AdS black hole with hyperbolic horizon $\Lambda=-\frac{3}{l^2}$ and $\gamma=-1$ \cite{Martinez:2005di}. However, one may note the other two case correspond to no black hole but solutions with naked singularity.
Then the metric function $f(r)$ (\ref{f(r)}) takes the form
\begin{align}
   f(r)=\mp\frac{r^2}{l^2}\pm\left(1\mp\frac Mr\right)^2
\end{align}
where the upper and lower of sign stand for the dS and AdS solution respectively.
There are four roots for this metric function \cite{Barlow:2005yd,Mann:1997jb}
\begin{align*}
r_1 &=\frac l2\left(1+\sqrt{1\mp\frac{4M}{l}}\right), &
r_2 &=\frac l2\left(1-\sqrt{1\mp\frac{4M}{l}}\right),\\
r_3 &=\frac l2\left(-1+\sqrt{1\pm\frac{4M}{l}}\right), &
r_4 &=\frac l2\left(-1-\sqrt{1\pm\frac{4M}{l}}\right).
\end{align*}
Both black holes have possessed cosmological, event and inner horizons, given by the radial coordinate
as $r_1$, $r_2$, $r_3$ respectively, and the $r_4$ is corresponding to a ``virtual'' horizon. One need note that, in fact we are considering the black holes with some special black hole mass $M$, in order to have multi-real roots, as we are interested in the ''entropy product'' of multi-horizons black hole.

The entropy corresponds to each horizon is
\begin{equation}
S(r_i)=\pi r_i^2\left(1-\frac{\phi(r_i)^2}{6}\right).
\end{equation}
where the scalar field behaviours as \cite{Martinez:2005di}
\begin{equation}
  \phi(r)=\sqrt{-\frac{\Lambda}{6\alpha}}\left(\frac{M}{r\mp M}\right)=\sqrt{\mp\frac{1}{2\alpha
l^2}}\left(\frac{M}{r\mp M}\right)
\end{equation}
Note $\alpha$ and $\Lambda$ have opposite signs.
After some direct calculation, we can find ''entropy sum'' of all four horizons
\begin{equation}
\sum_{i=1}^4 S(r_i)=2\pi l^2\pm\frac{\pi}{6\alpha}=\pm\frac{6\pi}{\Lambda}\pm\frac{\pi}{6\alpha}
\end{equation}
To give the conclusion, when there is extra scalar field in the spacetime, one will find the ``entropy sum''  is dependent of the cosmological constant and the constant signifying the coupling strength between gravity and the scalar field in the conformally coupling frame. It is also interesting to consider it in the minimally coupling frame. The result is shown in appendix \hyperref[appsec:Conformal transformation to change the non-minimal coupled to minimal coupled]{A}. We find the``entropy sum'' is also dependent of the cosmological constant and the constant characterizing the strength of self-interacting potential of the scalar field. What we emphasize is the ``virtual'' horizon cannot be dropped, otherwise we cannot get the ``entropy sum''  result, which has background field constant dependence and conserved charge independence (Here is $M$-independence and $Q$-independence). In this sense, we say the ``entropy sum'' is ``universal'' in this theory. On the other hand, comparing this case with that of kerr-(A)dS black hole (see Eq.(\ref{kerr})), it seems like that the topology of the sub-manifold $\Sigma$ can modify the ``entropy sum'' in someway.

\subsection{Charged rotating and static black holes in Einstein-Weyl theory}
\label{subsec:Charged rotating and static black holes in Einstein-Weyl theory}
The dyonic black hole solution in $D=4$ charged
Einstein-Weyl theory has the Lagrangian \cite{Liu:2012xn,Cvetic:2013eda}.
\begin{align}
\mathcal{L}&=\sqrt{-g}\left(\frac 12\alpha C^{\mu\nu\rho\sigma}C_{\mu\nu\rho\sigma}+\frac 13\alpha
            F^{\mu\nu}F_{\mu\nu}\right) \notag \\
           &=\sqrt{-g}\left(\alpha R^{\mu\nu}R_{\mu\nu}-\frac 13\alpha R^2+\frac 13\alpha F^2\right)
            +\alpha\mathcal{L}_{GB}.
\end{align}
where $\mathcal{L}_{GB}$ denotes the Gauss-Bonnet Lagrangian.
And the charged rotating AdS black hole solution can be written as \cite{Liu:2012xn,Cvetic:2013eda}
\begin{align}
ds_4^2=\rho^2\left(\frac{dr^2}{\Delta_r}+\frac{d\theta^2}{\Delta_\theta}\right)+
       \frac{\Delta_\theta \sin^2\theta}{\rho^2}\left(adt-(r^2+a^2)\frac{d\phi}{\Xi}\right)^2-
       \frac{\Delta_r}{\rho^2}\left(dt-a\sin^2\theta\frac{d\phi}{\Xi}\right)^2,
\end{align}
where
\begin{align*}
\rho^2&=r^2+a^2\cos^2\theta, &
\Delta_\theta&=1-g^2a^2\cos^2\theta,\\
\Xi&=1-g^2a^2, &
\Delta_r&=(r^2+a^2)(1+g^2r^2)-2mr+\frac{(p^2+q^2)r^3}{6m},
\end{align*}
where $\Lambda=-3g^2$ is the cosmological constant. In what follows we have set magnetic charge $p=0$ so that there
is only an electric charge $q$. Solve the equation $\Delta_r=0$, one can obtain four horizons. The entropy on these horizons are Wald entropy which do not satisfy the area theorem, derived in
\cite{Liu:2012xn} and have the form \cite{Liu:2012xn,Cvetic:2013eda}.
\begin{equation}
S(r_i)=\frac{2\pi \alpha}{\Xi}\left(1+g^2 r_i^2+\frac{q^2\,r_i}{6m}-c\Xi\right),
\end{equation}
where the constant $c$ is numerical and corresponds to adding
a constant multiple of the Gauss-Bonnet invariant to the action
\cite{Cvetic:2013eda}. Directly we calculate the sum of horizons entropy
\begin{equation} \label{eq:entropy sum of Charged rotating BH}
\sum_{i=1}^4 S(r_i)=4\pi\alpha(1-2c)
\end{equation}
This result does not relate to the conserved quantities: total energy $E$, charge $Q$ and angular
momentum $J$ \cite{Liu:2012xn,Cvetic:2013eda}.
\begin{align*}
E&=\frac{2\alpha g^2}{\Xi^2}\left(m+\frac{a^2q^2}{12m}\right)\\
Q&=\frac{\alpha q}{3\Xi}\\
J&=\frac{2a\alpha g^2}{\Xi^2}\left(m+\frac{q^2}{12m g^2}\right)
\end{align*}
Thus, the ``entropy sum''  is shown a only dependence of the constant $\alpha$, even the cosmological constant dependence is vanishing.

Next we consider a special case, the charged static dS black hole solution i.e. $J=0$, which the four horizons reduce to three \cite{Cvetic:2013eda}. Here $\Lambda=3g^2$ is the cosmological constant. The metric of static black hole is (a detail analysis of general solution is given in \cite{Riegert:1984zz})
\begin{align}
&ds^2=-f\,dt^2+\frac{dr^2}{f}+r^2d\Omega^2_2  \notag \\
&A=-\frac{q}{r}dt\\
&f=-\frac{\Lambda r^2}{3}+c_1r+c_0+\frac dr\\
&3c_1d+1+q^2=c_0^2.  \notag
\end{align}
For the static black hole, the entropy for each horizon is \cite{Cvetic:2013eda}
\begin{equation}
S(r_i)=-\frac{2\pi\alpha(3d+(c_0+2)r_i)}{3r_i}.
\end{equation}
We calculate the ``entropy sum'' to be
\begin{equation} \label{eq:entropy sum of Charged static BH}
\sum_{i=1}^3 S(r_i)=-4\pi\alpha.
\end{equation}

There are no conserved quantities $E, Q$
in the ``entropy sum''  (\ref{eq:entropy sum of Charged rotating BH}) and (\ref{eq:entropy sum of
Charged static BH}). So we could say that the ``entropy sum''  is shown a only dependence of the constant $\alpha$, which characterizes the strength of higher curvature terms, even the cosmological constant dependence is vanishing.

\section{Conclusion and Discussion}
In this paper, we find another ``universal property''  of entropy,
the``entropy sum'' relation of black holes in four dimensional
(anti-)de-Sitter asymptotical background. We mainly test ``entropy
sum'' relation in static, stationary black hole and some black
holes with other extra matter source (scalar hair and higher
curvature terms) in the asymptotical (anti-)de-sitter spacetime
background. They depend only on the cosmological constant with the
necessary effect of the un-physical ``virtual'' horizon included
and in the spacetime only cosmological constant, mass of the black
hole, rotation parameter and the Maxwell field exist. When there
is more extra matter field in the spacetime, one will find the
``entropy sum'' is also dependent of the strength of these extra
matter field. For both cases, we conclude that the new ``universal
property'', that is, the``entropy sum''  does not depend on the
conserved charges: $M$ (mass), $Q$ (from Maxwell field) and $J$
(from rotation), while it does depend on the property of
background spacetime. To say it more accurately, it does depend on
those constants, which characterize the strength of the background
fields. When there is extra degree of freedom, that is the scalar field in
the spacetime, it is dependent on the cosmological constant and
the constant signifying the strength of self-interacting potential of the scalar field in both the conformally coupling frame and in the
minimally coupling frame as shown in appendix
\hyperref[appsec:Conformal transformation to change the non-minimal coupled to minimal coupled]{A}. Besides, in
the Einstein-Maxwell-scalar-AdS spacetime, it seems like that the
topology of the sub-manifold $\Sigma$ can modify the ``entropy
sum'' in someway; we also point out the mass independence of the
''entropy product'' relation failed in this case. When we consider
the charged rotating and static black holes in the Einstein-Weyl
theory, the ``entropy sum''  is shown to be only dependence on the
constant characterizing the strength of higher curvature terms,
even if the cosmological constant dependence is vanishing. What we
emphasize is the ``virtual'' horizon cannot be dropped, otherwise
we cannot get the ``entropy sum'' relation with the background
field constant dependence. In this sense, we say the ``entropy
sum'' is ``universal'' in the theory presented in this paper. One
shall note that the ``entropy sum'' is negative in some black hole
case, which maybe result from the effect of the entropy of the
work out ``virtual'' horizon.

To give a whole look of the ``entropy sum'', we finally consider it
in the Kerr black hole case as shown in appendix
\hyperref[appsec:Kerr solution in flat]{B}. We find only mass
dependence of ``entropy sum'' appears. It makes us believe that,
considering the dependence properties of ``entropy sum'' relation,
the mass background field may be regarded as the next order of
cosmological constant background field and extra matter field,
while the Maxwell field and ``rotation field'' always play no role.
Explaining the relationship between the``entropy sum'' and
background properties still are open problems, which is left to be
a future work.

\section*{Acknowledgments}
We thank Liu Zhao and Bin Wu for useful comments and enlightening
discussions.This work is partially supported by the Natural
Science Foundation of China (NSFC) under Grant No.11075078.

\appendix
\section{The ``entropy sum''  in the minimally coupling frame of Einstein-Maxwell-scalar-AdS spacetime}
\label{appsec:Conformal transformation to change the non-minimal coupled to minimal coupled}
We consider the ``entropy sum'' in the minimally coupling frame of Einstein-Maxwell-scalar-AdS spacetime. One can obtain the solution directly by taking a conformal transformation of that in the conformally coupling frame \cite{Martinez:2005di,Martinez:2004nb,Maeda:1988ab}. Here we take the AdS black hole with hyperbolic horizon $\Lambda=-\frac{3}{l^2}$ and $\gamma=-1$ \cite{Martinez:2005di} in Section \ref{subsec:Scalar Hairy Black Holes} as an example. The corresponding conformal transformation is
\begin{equation}  \label{eq:conformal factor}
\Omega^2=1-\frac 16\phi^2=1-\frac{1}{12\alpha l^2}\left(\frac{M}{r+M}\right)^2.
\end{equation}
Then we introduce a new scalar field $\Phi$
\begin{equation}
\tanh\left(\sqrt{\frac 16}\Phi\right)=\sqrt{\frac 16}\phi
\end{equation}
with $\Phi(r)$ behaviour as
\begin{equation}
\Phi(r)=\sqrt{6}\operatorname{arctanh}\left(\sqrt{\frac{1}{12\alpha
l^2}}\frac{M}{r+M}\right)
\end{equation}
One can obtain the theory in the minimally coupling frame with the action
\begin{align}
\mathcal{L}=\int d^4x\sqrt{-\hat{g}}\left(\frac{\hat{R}}{16\pi}-\frac 12
\hat{g}^{\mu\nu}\partial_{\mu}\Phi\partial_{\nu}\Phi-V(\Phi)\right)-\frac{1}{16\pi}\int
d^4x\sqrt{-\hat{g}}F^{\mu\nu}F_{\mu\nu},
\end{align}
where the new self-interaction potential $V(\Phi)$ takes the form
\begin{equation}
V(\Phi)=\frac{\Lambda}{8\pi}\left(\cosh^4\sqrt{{\frac 16}}\Phi
        +\frac{8\pi}{\Lambda}\,36\alpha \sinh^4\sqrt{\frac 16}\Phi\right)
\end{equation}
The transformed, minimal coupled version line element is
\begin{align}
d\hat{s}^2=\Omega^2\left[-\left(\frac{r^2}{l^2}-\left(1+\frac{M}{r}\right)^2\right)dt^2+
\frac{dr^2}{\frac{r^2}{l^2}-\left(1+\frac{M}{r}\right)^2}+r^2 d\sigma^2\right]
\end{align}
We still need to introduce a new radial coordinate $R^2=r^2\Omega^2$ to get the familiar coordinate frame, which one will obtain the usual Benkenstain-Hawking area entropy
\begin{equation}
S(R_i)=\frac{A(R_i)}{4}=\frac{\sigma R_i^2}{4}
\end{equation}
Then some calculation lead to
\begin{align*}
  \sum_{i=1}^4S_i&=\sum_{i=1}^4S(R_i)=\sum_{i=1}^4\frac{\sigma r_i^2}{4}\Omega^2\\
  &=\sum_{i=1}^4\pi r_i^2\left(1-\frac{\phi(r_i)^2}{6}\right)\\
  &=-\frac{6\pi}{\Lambda}-\frac{\pi}{6\alpha}
\end{align*}
which is the same as shown in Section \ref{subsec:Scalar Hairy Black Holes}. Obviously the same rules of
``entropy sum''  still holds. When in the minimally coupling frame, ``entropy sum'' is also dependent of the cosmological constant and the constant characterizing the strength of self-interacting potential of the scalar field.

\section{The``entropy sum'' of Kerr black hole} \label{appsec:Kerr solution in flat}
To give a whole look of the ``entropy sum'', we consider it in asymptotical flat spacetime without any matter source, taking Kerr black hole \cite{Kerr:1963ud} as an example. All horizons of Kerr black hole are
\begin{align*}
r_1=M+\sqrt{M^2-a^2}\\
r_2=M-\sqrt{M^2-a^2}
\end{align*}
and area of each horizon are
\begin{align}
A(r_i)=4 \pi (r_i^2+a^2) \notag \\
\intertext{entropy respectively}
S(r_i)=\frac{A(r_i)}{4}
\end{align}
then ``entropy sum'' is
\begin{equation}
\sum_{i=1}^2S(r_i)=4\pi M^2
\end{equation}
apparently it is mass dependent. It seems that, considering the dependence of ``entropy sum'' , the mass background field maybe is the next order of cosmological constant background field and extra matter field, while the Maxwell field and ``rotation field'' always play no role.

\providecommand{\href}[2]{#2}\begingroup
\footnotesize\itemsep=0pt
\providecommand{\eprint}[2][]{\href{http://arxiv.org/abs/#2}{arXiv:#2}}

\end{document}